\documentclass[elsart12,epsf,eqsecnum]{revtex4}
\usepackage{amsmath}
\usepackage{epsfig,epsf}
\usepackage{graphicx}
\usepackage{verbatim}
\usepackage{epsfig,epsf}
\begin{document}

\newcommand{\W} {\text{W}}
\def\beq{\begin{equation}}
\def\eeq{\end{equation}}

\def\eq#1{{Eq.~(\ref{#1})}}
\def\fig#1{{Fig.~\ref{#1}}}


\title{Remarks on Diffractive Dissociation  within JIMWLK Evolution at NLO.}
\author{Michael Lublinsky}

\affiliation{Physics Department, Ben-Gurion University of the Negev, Beer Sheva 84105, Israel
 }
\begin{abstract}
We discuss the high energy diffractive dissociation in DIS at the  Next to Leading Order. In the large $N_c$ dipole limit 
 we derive the NLO version of the Kovchegov-Levin equation. We argue that the original structure of the equation is preserved, 
 that is it coincides with the Balitsky-Kovchegov equation at NLO.
\end{abstract}


\maketitle


\mbox{}

\pagestyle{plain}

\setcounter{page}{1}




\date{\today}

\section{Introduction and Conclusion}

In recent years a lot of attention has been devoted to development and phenomenological applications of the theory of perturbative saturation \cite{GLR}. The theoretical description of the energy evolution of the wave function towards  a dense state at leading order in $\alpha_s$ has been long known.
 It is given by  JIMWLK equation \cite{cgc,jimwlk}, or equivalently Balitsky hierarchy \cite{Bal}.
 It generalizes the well known BFKL equation \cite{bfkl}  by including finite density effects in the hadronic wave function. 
 JIMWLK   is restricted  in applicability  to DIS-type processes when target is large while projectile is small throughout the entire range in  rapidity $Y$ of the process. 
The mean field approximation to the JIMWLK equation, the  Balitsky-Kovchegov (BK) equation \cite{Bal,KOV} has been used extensively during the last decade in numerous phenomenological applications, including fits to DIS low x data and the diffraction   measured by HERA. 
The latter process is frequently regarded as a direct measure of non-linear high density effects.
For JIMWLK-based phenomenological applications it is important to include perturbative corrections beyond leading order, since 
they are known to lead to large effects already in the linear BFKL framework \cite{NLOBFKL}.

Based on the milestone results of \cite{BC} and \cite{Grab},  in a recent paper \cite{nlojimwlk} we have derived the complete operator form of the JIMWLK Hamiltonian at the
next to leading order. This paper appeared simultaneously with \cite{BClast}, which directly calculated the elements of the general Balitsky hierarchy at NLO.  In \cite{KLMconf}  
we show that the NLO JIMLWK equation for ${\cal N}=4$ theory has exact conformal invariance, even though it is derived with sharp rapidity cutoff.

At leading order,   diffraction in saturation environment was explored  by  Kovchegov and Levin \cite{Kovchegov:1999ji}.
The diffractive cross section $N^D_{El}$ investigated  in
\cite{Kovchegov:1999ji}  is diffractive on the projectile side but  elastic in the target degrees freedom.
It was shown in \cite{Kovchegov:1999ji}  that  in the dipole limit, $N^D_{El}$ obeys the  BK equation. 
Running coupling effects \cite{weigertrun} for the process have been included in  \cite{Kovrundiff}. 

The aim of the present note is to  address diffractive dissociation  within the NLO JIMWLK.  In \cite{KLW} we  developed the formalism
for calculating diffractive dissociation and other semi-inclusive observables within the JIMWLK framework. We
considered  several examples, in particular variety of elastic and/or diffractive
cross sections, cross section with fixed transverse momentum transfer and
inclusive gluon spectrum. In all these cases we defined the appropriate
observable, derived its evolution with rapidity (total rapidity of the process
and/or width of the rapidity gap and/or width of the diffractive interval) and
discussed  the dipole model limit for each one of the observables.
Our main observation here is that many results  of \cite{KLW}  are in fact  independent of the
explicit form of the JIMWLK Hamiltonian and thus generalize straightforwardly to  the NLO accuracy.


The following two points summarize our findings. First, the expression for diffractive dissociation (\ref{PD1})  which was originally derived in \cite{KLW} 
is  valid as long as the high energy factorization between projectile and target holds. Particularly it is valid,  when the projectile is dilute, target is dense, 
and the respective evolution is governed by the NLO JIMWLK.  The second point  is  that, just like at LO,  the large $N_c$
dipole Hamiltonian at NLO is linear in the dipole conjugate field.  This makes the NLO version of the Kovchegov-Levin 
to coincide with the BK equation at NLO.  In order to compute  diffractive DIS cross
sections, one needs to convolute $N^D_{El}$ with the photon impact  factor \cite{BCif}.

\section{High energy scattering}

In this section we recap some basic formalism and introduce notations. 
The total $S$-matrix of the high energy scattering process at a given rapidity
$Y$ in the CGC formalism is computable via the following factorization formula
\begin{equation}
{\cal S}(Y)\,=\,\int\, DS\,\, 
W^T_{Y_0}[S]\,\,\Sigma^{PP}_{Y-Y_0}[S]\,.
\label{ss}
\end{equation}
Here for a composite projectile which has some distribution of gluons in its wave function
 the eikonal $S$-matrix is
\begin{equation}
\Sigma^{PP}_{Y-Y_0}[S]\,\equiv\,\langle P|\,\hat S\,|P\rangle\,=\,\int d\rho\,\,W^P_{Y-Y_0}[\rho]\,\,
\,\exp\left\{i\int d^2x\,\rho^a(x)\,\alpha^a(x)\right\}
\label{s1}
\end{equation}
 $\rho(x)$ is the color charge density in the projectile wave function at a given transverse
position;  $W^P_{Y-Y_0}[\rho]$ is its probability distribution in the projectile, while $\alpha$ is a target color field.
In eq.(\ref{ss}), the projectile-averaged $S$-matrix  $\Sigma^{PP}$  is further averaged over the distribution of the
color fields $\alpha$ with the weight $W^T_{Y_0}[S(x)]$. 
The fields $\alpha$   are parametrized by the eikonal $S$-matrix $S(x$) for a single parton at transverse
position $x$ to scatter on a given configuration of $\alpha$. 

In \eq{ss} we have chosen the frame where the target has rapidity $Y_0$ while
the projectile carries the rest of the total rapidity $Y-Y_0$.  Lorentz invariance requires ${\cal S}$ to be independent of $Y_0$.
The high energy evolution of both $W^{P.T}$ is driven by an effective high energy Hamiltonian,
 which in this note will be assumed to be the NLO JIMWLK 
\begin{equation}
\frac{d}{d\,Y}\,{W_Y^{P,T}}\,=-\,H^{NLO\,JIMWLK}\ W_Y^{P,T}\label{hee}
\end{equation}

The NLO JIMWLK Hamiltonian \cite{nlojimwlk} is 
\begin{eqnarray}\label{NLO}
H^{NLO\ JIMWLK}&=& \int_{x,y,z}K_{JSJ}(x,y;z) \left[ J^a_L(x)J^a_L(y)+J_R^a(x)J_R^a(y)-2J_L^a(x)S_A^{ab}(z)J^b_R(y)\right] + \nonumber \\
 &&+\int_{x\,y\, z\,z^\prime} K_{JSSJ}(x,y;z,z^\prime)\left[f^{abc}f^{def}J_L^a(x) S^{be}_A(z)S^{cf}_A(z^\prime)J_R^d(y)- N_c J_L^a(x)S^{ab}_A(z)J^b_R(y)\right] 
 +\nonumber \\
 &&+\int_{x,y, z,z^\prime} K_{q\bar q}(x,y;z,z^\prime)\left[2\,J_L^a(x) \,tr[S^\dagger(z)\, T^a\,S (z^\prime)T^b]\,J_R^b(y)\,
 -\, J_L^a(x)\,S^{ab}_A(z)\,J^b_R(y)\right]+\nonumber \\
 &&+\int_{w,x,y, z,z^\prime}K_{JJSSJ}(w;x,y;z,z^\prime)f^{acb}\,\Big[J_L^d(x)\, J_L^e(y)\, S^{dc}_A(z)\,S^{eb}_A(z^\prime)\,J_R^a(w)\,-\nonumber \\
 &&-J_L^a(w)\,S^{cd}_A(z)\,S^{be}_A (z^\prime)\,J_R^d(x)\,J_R^e(y)\,+\frac{1}{3}[\,J_L^c(x) \,J_L^b(y) \,J_L^a(w)\,-\, 
J_R^c(x) \,J_R^b(y)\,J_R^a(w)]\,\Big] \,
 +\nonumber \\
 &&+\int_{w,x,y, z}\, K_{JJSJ}(w;x,y;z)\,f^{bde}\,\Big[  J_L^d(x) \,J_L^e(y) \,S^{ba}_A(z)\,J_R^a(w)\,-\, 
J_L^a(w)\,S^{ab}_A(z)\,J_R^d(x)\,J_R^e(y) \, +\, \nonumber \\
 &&+\frac{1}{3}[\,J_L^d(x) \,J_L^e(y) \,J_L^b(w)\,-\, 
J_R^d(x)\,J_R^e(y)\,J_R^b(w)]
\Big] 
\end{eqnarray}
Here $S_A$ is a unitary matrix in the adjoint representation - the gluon scattering amplitude.
The left and right $SU(N_c)$ rotation generators, when acting on functions of $S$ have the representation
\begin{eqnarray}\label{LR}
J^a_L(x)=tr\left[\frac{\delta}{\delta S^{T}_x}T^aS_x\right]-tr\left[\frac{\delta}{\delta S^{*}_x}S^\dagger_xT^a\right] ; \ \ 
J^a_R(x)=tr\left[\frac{\delta}{\delta S^{T}_x}S_xT^a\right] -tr\left[\frac{\delta}{\delta S^{*}_x}T^aS^\dagger_x\right]\,.
\end{eqnarray}
Here $T^a$ are $SU(N_c)$ generators in the fundamental representation.
We use the notations of ref. \cite{BC} $X\equiv x-z$,  $X^\prime\equiv x-z^\prime$, $Y\equiv y-z$,    $ Y^\prime \equiv y-z^\prime$,
$W\equiv w-z$,    $ W^\prime \equiv w-z^\prime$, and $Z\equiv z-z^\prime$. 
All $J$s in  (\ref{NLO}) are assumed not to act on $S$ in the Hamiltonian.  
\begin{eqnarray}
K_{JSJ}(x,y;z) =-\frac{\alpha_s^2}{16 \pi^3}
\frac{(x-y)^2}{X^2 Y^2}\Big[b\ln(x-y)^2\mu^2
-b\frac{X^2-Y^2}{ (x-y)^2}\ln\frac{X^2}{Y^2}+
(\frac{67}{9}-\frac{\pi^2}{ 3})N_c-\frac{10}{ 9}n_f\Big]\,
-\, \frac{N_c}{2}\ \int_{z^\prime}\, \tilde K(x,y,z,z^\prime)\nonumber \\
\end{eqnarray}
Here $\mu$ is the normalization point in the $\overline{MS}$ scheme and
$b=\frac{11}{3}N_c-\frac{2}{3}n_f$. \begin{eqnarray}
&&K_{JSSJ}(x,y;z,z^\prime) \ =\ \frac{\alpha_s^2}{16\,\pi^4}
\Bigg[\,-\,\frac{4}{ Z^4}\,+\,
\Big\{2\frac{X^2{Y'}^2+{X'}^2Y^2-4(x-y)^2Z^2}{ Z^4[X^2{Y'}^2-{X'}^2Y^2]}\nonumber\\ 
&&+
~\frac{(x-y)^4}{ X^2{Y'}^2-{X'}^2Y^2}\Big[
\frac{1}{X^2{Y'}^2}+\frac{1}{ Y^2{X'}^2}\Big]
+\frac{(x-y)^2}{Z^2}\Big[\frac{1}{X^2{Y'}^2}-\frac{1}{ {X'}^2Y^2}\Big]\Big\}
\ln\frac{X^2{Y'}^2}{ {X'}^2Y^2}\Bigg]\,+\tilde K(x,y,z,z^\prime) \end{eqnarray}
\begin{eqnarray}
\tilde K(x,y,z,z^\prime)=\frac{i}{2}\,\left[K_{JJSSJ}(x;x,y;z,z^\prime)-K_{JJSSJ}(y;x,y;z,z^\prime)\,-\,
K_{JJSSJ}(x;y,x;z,z^\prime)+K_{JJSSJ}(y;y,x;z,z^\prime)\right]
\end{eqnarray}
\begin{eqnarray}
K_{q\bar q}(x,y;z,z^\prime) =-\frac{\alpha_s^2\,n_f}{ 8\,\pi^4}
\Big\{
\frac{{X'}^2Y^2+{Y'}^2X^2-(x-y)^2Z^2}{ Z^4(X^2{Y'}^2-{X'}^2Y^2)}
\ln\frac{X^2{Y'}^2}{ {X'}^2Y^2}-\frac{2}{Z^4}\Big\}
\end{eqnarray}
\begin{eqnarray}
&&K_{JJSJ}(w;x,y;z)\,=\,-\,i\,\frac{\alpha_s^2}{ 4\, \pi^3 }\,\Big[ \frac{X\cdot W}{ X^2\,W^2}\,-\, \frac{Y\cdot W}{ Y^2\,W^2}    \Big] \ln\frac{Y^2}{ (x-y)^2}\,\ln\frac{X^2}{ (x-y)^2}\\
&&K_{JJSSJ}(w;x,y;z,z^\prime)=-i
\frac{\alpha_s^2}{ 2\,\pi^4}
\left(\frac{X_iY^\prime_j}{ X^2Y^{\prime 2}}
\right)\Big(\frac{\delta_{ij}}{2 Z^2}-\frac{Z_i W^\prime_j}{ Z^2 W^{\prime 2}}+
\frac{Z_j W_i}{ Z^2 W^{ 2}}-\frac{W_i W^\prime_j}{W^2 W^{\prime 2}}
\Big)\ln\frac{W^2}{ {W'}^2} \nonumber \\ 
\end{eqnarray}

Our goal will be to study diffractive dissociation at NLO within the dipole model approximation.
The $S$-matrix for a quark-antiquark dipole is
\begin{equation} 
   s(x,y)= {1\over N_c}tr[S_F(x)\, S^\dagger_F(y)] 
\end{equation}
where $F$ denotes fundamental representation. 
The $S$ matrix of a projectile made of several but not too many  dipoles 
is therefore some function of the variable $s$ only 
\begin{equation}
\Sigma^{PP}[S]=\Sigma^{PP}[s] \,.
\end{equation}
In the large $N_c$ limit, the NLO JIMWLK Hamiltonian acting on a function of dipoles only,  reduces to the action of the NLO dipole Hamiltonian.
To derive this Hamiltonian, we have to act with the NLO JIMWLK Hamiltonian on one, two and three dipoles. Four and more dipoles cannot be coupled by the NLO evolution because there are only three $J$s in the Hamiltonian.   
The action on a single dipole is by construction \cite{nlojimwlk}
reproduces the result of \cite{BC}
and, in the large $N_c$ limit  is equivalent to the action of the dipole Hamiltonian 
\begin{eqnarray}\label{NLOdipole}
H^{NLO\,dipole}&=&\frac{\alpha_{s}N_{c}}{2\pi^{2}}\int_{u,v}\Bigg[\int d^{2}z\frac{(u-v)^{2}}{U^{2}V^{2}}\left\{ 1+\frac{\alpha_{s}}{4\pi}\left[b\ln(u-v)^{2}\mu^{2}-b\frac{U^{2}-V^{2}}{(u-v)^{2}}\ln\frac{U^{2}}{V^{2}}+\left(\frac{67}{9}-\frac{\pi^{3}}{3}\right)N_{c}-\right.\right.\nonumber \\
&-&\left.\left.\frac{10}{9}n_{f}-2N_{c}\ln\frac{U^{2}}{(u-v)^{2}}\ln\frac{V^{2}}{(u-v)^{2}}\right]\right\} \left[s(u,z)s(z,v)-s(u,v)\right]+\nonumber \\
&+&\frac{\alpha_{s}^{2}}{16N_{c}\pi^{4}}\int d^{2}zd^{2}z^{\prime}\Big[\left(-\frac{4}{(z-z^{\prime})^{4}}+\left\{ 2\frac{U^{2}(V^{\prime})^{2}+(U^{\prime})^{2}V^{2}-4(u-v)^{2}(z-z^{\prime})^{2}}{(z-z^{\prime})^{2}\left[U^{2}(V^{\prime})^{2}-(U^{\prime})^{2}V^{2}\right]}+\right.\right.\nonumber \\
&+&\left.\left.\frac{(u-v)^{4}}{U^{2}(V^{\prime})^{2}-(U^{\prime})^{2}V^{2}}\left[\frac{1}{U^{2}(V^{\prime})^{2}}+\frac{1}{V^{2}(U^{\prime})^{2}}\right]+\frac{(u-v)^{2}}{(z-z^{\prime})^{2}}\left[\frac{1}{U^{2}(V^{\prime})^{2}}-\frac{1}{(U^{\prime})^{2}V^{2}}\right]\right\} \ln\frac{U^{2}(V^{\prime})^{2}}{(U^{\prime})^{2}V^{2}}\right)\times\nonumber \\
&\times&\left[N_{c}^{3}s(u,z^{\prime})s(z^{\prime},z)s(z,v) -(z^{\prime}\rightarrow z)\right]\nonumber \\
&-&\left\{ \frac{(u-v)^{2}}{(z-z^{\prime})^{2}}\left[\frac{1}{U^{2}(V^{\prime})^{2}}+\frac{1}{V^{2}(U^{\prime})^{2}}\right]-\frac{(u-v)^{4}}{V^{2}(U^{\prime})^{2}(V^{\prime})^{2}U^{2}}\right\} \ln\left(\frac{U^{2}(V^{\prime})^{2}}{(U^{\prime})^{2}V^{2}}\right)N_{c}^{3}s(u,z^{\prime})s(z^{\prime},z)s(z,v)+\nonumber \\
&+&N_c\,n_{f}\left\{ \frac{4}{(z-z^{\prime})^{4}}-2\frac{U^{2}(V^{\prime})^{2}+(U^{\prime})^{2}V^{2}-(u-v)^{2}(z-z^{\prime})^{2}}{(z-z^{\prime})^{4}((V^{\prime})^{2}U^{2}-V^{2}(U^{\prime})^{2})}\ln\left(\frac{U^{2}(V^{\prime})^{2}}{(U^{\prime})^{2}V^{2}}\right)\right\} \,s(u, z^\prime)s(z,v)\Big]\Bigg] \frac{\delta}{\delta s(u, v)} \nonumber \\
\end{eqnarray}
The connected parts of two- and three-dipole evolutions are subleading in $N_c$ and  it is a matter of a straightforward color algebra to see that. When acting with the NLO JIMWLK  on two dipoles, the connected part arises when both dipoles are rotated by at least one $J$. The $N_c$ counting of such a "dipole merging" is most easily done when $S_A(z)$ and $S_A(z^\prime)$ in the Hamiltonian are set to  one. All the "dipole merging" terms generated by the operators  in the NLO JIMWLK Hamiltonian are found to be subleading in $N_c$ compared to uncorrelated terms generated by $ H^{NLO\,dipole}$.  
 As a result, there are no leading $N_c$  ${\delta^2}/{\delta s^2}$ and   (${\delta^3}/{\delta s^3}$ terms in the dipole Hamiltonian.


Similarly to the LO case, the Hamiltonian $H^{\text{NLO\, dipole}}$ is linear with respect to the dipole conjugate field $\frac{\delta}{\delta s(u, v)}$. 
Thanks to this  important property,  the dipole evolution equation
\begin{equation}
\label{dipoleev}
{d\Sigma_Y^{PP}[s]\over dY}\,=-\,H^{\text{dipole}}\,\Sigma^{PP}_Y[s]
\end{equation}
can be solved as 
\begin{equation}\label{sig}
\Sigma_Y^{PP}[s]\,=\,\Sigma_{Y_0}^{PP}[s_Y] 
\end{equation}
where $s_Y$ solves the  NLO BK equation
\begin{equation}\label{NLOBK}
\begin{split}\frac{d}{dY}s(u,v)&=\frac{\alpha_{s}N_{c}}{2\pi^{2}}\int d^{2}z\frac{(u-v)^{2}}{U^{2}V^{2}}\left\{ 1+\frac{\alpha_{s}}{4\pi}\left[b\ln(u-v)^{2}\mu^{2}-b\frac{U^{2}-V^{2}}{(u-v)^{2}}\ln\frac{U^{2}}{V^{2}}+\left(\frac{67}{9}-\frac{\pi^{3}}{3}\right)N_{c}-\frac{10}{9}n_{f}-\right.\right.\\
&\left.\left.-2N_{c}\ln\frac{U^{2}}{(u-v)^{2}}\ln\frac{V^{2}}{(u-v)^{2}}\right]\right\} \left[s(u,z)s(z,v)-s(u,v)\right]+\\
&+\frac{\alpha_{s}^{2}}{16N_{c}\pi^{4}}\int d^{2}zd^{2}z^{\prime}\left[\left(-\frac{4}{(z-z^{\prime})^{4}}+\left\{ 2\frac{U^{2}(V^{\prime})^{2}+(U^{\prime})^{2}V^{2}-4(u-v)^{2}(z-z^{\prime})^{2}}{(z-z^{\prime})^{2}\left[U^{2}(V^{\prime})^{2}-(U^{\prime})^{2}V^{2}\right]}+\right.\right.\right.\\
&\left.\left.+\frac{(u-v)^{4}}{U^{2}(V^{\prime})^{2}-(U^{\prime})^{2}V^{2}}\left[\frac{1}{U^{2}(V^{\prime})^{2}}+\frac{1}{V^{2}(U^{\prime})^{2}}\right]+\frac{(u-v)^{2}}{(z-z^{\prime})^{2}}\left[\frac{1}{U^{2}(V^{\prime})^{2}}-\frac{1}{(U^{\prime})^{2}V^{2}}\right]\right\} \ln\frac{U^{2}(V^{\prime})^{2}}{(U^{\prime})^{2}V^{2}}\right)\times\\
&\left[N_{c}^{3}s(u,z^{\prime})s(z^{\prime},z)s(z,v) -(z^{\prime}\rightarrow z)\right]\\
&-\left\{ \frac{(u-v)^{2}}{(z-z^{\prime})^{2}}\left[\frac{1}{U^{2}(V^{\prime})^{2}}+\frac{1}{V^{2}(U^{\prime})^{2}}\right]-\frac{(u-v)^{4}}{V^{2}(U^{\prime})^{2}(V^{\prime})^{2}U^{2}}\right\} \ln\left(\frac{U^{2}(V^{\prime})^{2}}{(U^{\prime})^{2}V^{2}}\right)N_{c}^{3}s(u,z^{\prime})s(z^{\prime},z)s(z,v)+\\
&\left.+N_c\,n_{f}\left\{ \frac{4}{(z-z^{\prime})^{4}}-2\frac{U^{2}(V^{\prime})^{2}+(U^{\prime})^{2}V^{2}-(u-v)^{2}(z-z^{\prime})^{2}}{(z-z^{\prime})^{4}((V^{\prime})^{2}U^{2}-V^{2}(U^{\prime})^{2})}\ln\left(\frac{U^{2}(V^{\prime})^{2}}{(U^{\prime})^{2}V^{2}}\right)\right\} \,s(u, z^\prime)s(z,v) \right]
\\
\end{split}\end{equation}
with the initial condition
\begin{equation}
s_{Y_0}(x,y)\,=\,s(x,y)
\end{equation}
The target  average in eq.(\ref{ss}) still allows to accommodate nontrivial, non-factorized multi-s
correlators $\langle s(x_1,y_1)\cdots
s(x_n,y_n)\rangle_T$, which have been recently argued \cite{ddgjlv,KLreview} to be of relevance to various two-particle correlations,
such as the "ridge".

Further simplification is achieved if one assumes that the dipoles scatter on
the target independently. This amounts to factorization of the target averages of
the  dipole $s$-matrices
\begin{equation}
\langle s(x,y)\, s(u,v)\rangle_T\,=\, \langle s(x,y)\rangle_T\langle s(u,v)\rangle_T
\end{equation} 
With this assumption, one replaces the ensemble average over target fields 
with a fixed initial function $s_{Y_0}(x,y)$.  We 
refer to this factorization property as the target mean field approximation.
Within the target mean field approximation
\begin{equation}
\langle \Sigma^{PP}_0[ s_Y ]\rangle_T\,=\,\Sigma^{PP}_0[\langle s_Y\rangle_T ]
\ .
\end{equation}
As has been  stressed in the past,   this mean field approximation does not follow  from
the dipole model approximation for the evolution kernel eq.(\ref{dipoleev}),
but  is an additional assumption about the properties of the target.

\section{Diffractive  Dissociation}

We will be interested in identifying the evolution governing
the diffractive observables both with respect to the total rapidity of the
process $Y$ and  the rapidity gap.
We now consider processes where the projectile diffracts within  rapidity interval $Y_P$. This interval is not necessarily small, so this type of observable can be evolved independently over the total rapidity $Y$ and the width of the diffractive interval $Y_P$. The target can either scatter elastically or can in principle also diffract. 
We consider  here the process where the scattering on the target side is elastic with the minimal gap  $Y_{gap}=Y-Y_P$.

In ref. \cite{KLW}  we have developed a formalism that makes it possible to generalize  (\ref{ss})  for semi-inclusive processes. 
Particularly, for elastic/diffractive processes one has to introduce two independent $S$-matrix variables, $S$ and $\bar S$ for  
the amplitude and its conjugate.
For the observable in question the cross section reads
\beq\label{PD1}
N^{D,Y_P}_E\,=\,
\int DS\,D\bar S\,\left(1\,-\,\Sigma^{PP}_{Y_P}[S]\,-\,\Sigma^{PP}_{Y_P}[\bar S^\dagger]
\,+\, \Sigma^{PP}_{Y_P}[\bar S^\dagger\, S]\right)\,\,W^T_{Y_{gap}}[S]\,W^T_{Y_{gap}}[\bar S]\,.
\eeq
What we have here is the target evolved through the gap $Y_{gap}$ both in the amplitude and its conjugate, while the projectile is evolved  
inclusively though the diffractive interval $Y_P$.  
The derivation of this result did not rely on any explicit form of the evolution Hamiltonian
and is thus valid for the case of the NLO JIMWLK.

We would like now to project this result into the dipole picture at large $N_c$.
At leading order, this observable in the dipole limit has been discussed by Kovchegov and Levin \cite{Kovchegov:1999ji}. 
 In the dipole limit the evolution of $\Sigma^{PP}$ with respect to the diffractive interval at fixed $Y_{gap}$ is given by 
 eq. (\ref{sig}).  Within the gap, due to independent target averaging over  $S$ and $\bar S$ in eq. (\ref{PD1}),
 the "composite" dipole made of $\bar S^\dagger S$ factorizes into the product of  two 
\beq
{1\over N_c}\langle {\rm tr} [\bar S^\dagger(x)S(x)S^\dagger(y)\bar S(y)]\rangle_T\,=\,
\langle {1\over N_c}{\rm tr}[ S(x)S^\dagger(y)]\rangle_T\,\,\langle {1\over N_c}{\rm tr}[\bar S(y)\bar S^\dagger(x)]\rangle_T
\eeq
with each  dipole evolving according to its own NLO  BK equation (\ref{NLOBK}). Similarly to the LO case, we have
  \beq\label{FPD1} 
N^{D,Y_P}_E\,=\, 1\,-\,2 \,{\cal S}(Y) \,+\,\Sigma^{PP}_0[s^{el}_{Y_P,Y_{gap}}]
\eeq
where $s^{el}_{Y_P,Y_{gap}}$ is obtained by 
first solving the NLO BK equation with respect to $Y_{gap}$ with the initial condition
\beq
s_{Y_{gap}=0}(x,y)\,=\,s(x,y)
\eeq
and then evolving by $Y_P$ with the very same equation for $s^{el}$  with the initial condition
\beq 
s^{el}_{Y_P=0,Y_{gap}}\,=\,s_{Y_{gap}}^2\,.
\eeq
As has been already noticed in \cite{KLW},  the derivation in \cite{Kovchegov:1999ji} was originally  done for  a single dipole projectile, $\Sigma^{PP}[s]=s$.
Thanks to the fact that at both  LO and NLO,  all the dipoles evolve independently, 
\eq{FPD1} provides   generalization to a more complex projectile wave function.


\section*{Acknowledgments}

The author thanks Alex Kovner and  the Physics Department of the University of Connecticut for the hospitality.
 The research was supported by the  EU FP7 grant PIRG-GA-2009-256313; the 
 ISRAELI SCIENCE FOUNDATION grant \#87277111;  the  BSF grant \#2012124; 
 and the People Program (Marie Curie Actions) of the European Union's Seventh Framework Program FP7/2007-2013/ under REA
grant agreement \#318921.

\end{document}